\documentclass[sigconf]{acmart}
\renewcommand\footnotetextcopyrightpermission[1]{}
\renewcommand\footnotetextcopyrightpermission[1]{}

\setcopyright{none}

\usepackage{algorithm}
\usepackage[noend]{algpseudocode}

\usepackage{subcaption}
\usepackage{textcomp}
\usepackage{tabularx}
\usepackage{tabu}
\usepackage{longtable}
\usepackage{enumitem}
\usepackage{makecell}
\usepackage{multirow}
\usepackage{xspace}
\usepackage{url}
\usepackage{listings}

\usepackage{booktabs}
\usepackage{graphicx}
\makeatletter
\def\do@url@hyp{\do\-\do\_}
\makeatother
\usepackage{flushend}

\newcommand{\PHB}[1]{\noindent\textbf{#1}\hspace{.5em}} 
\newcommand{\PHM}[1]{\vspace{.2em}
\noindent\textbf{#1}\hspace{.5em}} 

\newcommand{\Q}[1]{\noindent\textbf{#1}\hspace{.5em}}

\graphicspath{{figures/}}

\definecolor{mygreen}{rgb}{0,0.6,0}
\definecolor{mygray}{rgb}{0.5,0.5,0.5}
\definecolor{mymauve}{rgb}{0.58,0,0.82}
\definecolor{myred}{rgb}{0.79,0.15,0.15}

\newcommand{\SysName}{\texttt{Proteus}\xspace}

\begin{document}

\title{Making Serverless Computing Extensible: \\ A Case Study of Serverless Data Analytics}

\author{
	{\rm Minchen Yu, Yinghao Ren, Jiamu Zhao, Jiaqi Li}
} 
\affiliation{%
 \institution{
 The Chinese University of Hong Kong, Shenzhen
 }
}

\begin{abstract}

Serverless computing has attracted a broad range of applications due to its ease of use and resource elasticity. 
However, developing serverless applications often poses a dilemma---relying on general-purpose serverless platforms can fall short of delivering satisfactory performance for complex workloads, whereas building application-specific serverless systems undermines the simplicity and generality.
In this paper, we propose an \emph{extensible} design principle for serverless computing.
We argue that a platform should enable developers to extend system behaviors for domain-specialized optimizations while retaining a shared, easy-to-use serverless environment.
We take data analytics as a representative serverless use case and realize this design principle in \SysName.
\SysName introduces a novel abstraction of decision workflows, allowing developers to customize control-plane behaviors for improved application performance.
Preliminary results show that \SysName's prototype effectively optimizes analytical query execution and supports fine-grained resource sharing across diverse applications.

\end{abstract}

\maketitle

\section{Introduction}

\begin{figure}
    \centering
    \includegraphics[width=0.42\textwidth]{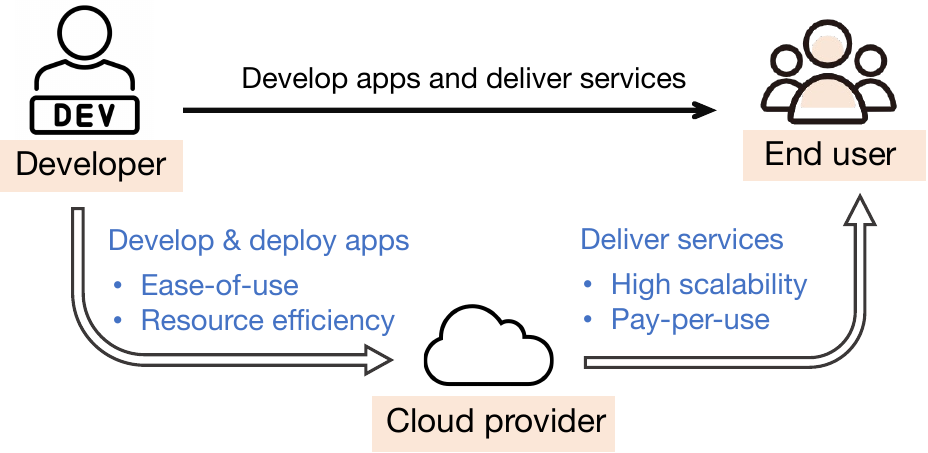}
    \caption{Serverless paradigm and the key stakeholders.}
    \label{fig:serverless}
    \vspace{-0.2in}
\end{figure}

Serverless computing, or Function-as-a-Service (FaaS), has emerged as a popular cloud computing paradigm and is widely adopted by major cloud providers, such as AWS Lambda~\cite{aws_lambda}, Google Cloud Functions~\cite{google_cloud_function}, and Azure Functions~\cite{azurefunc}. 
As illustrated in Fig.\ref{fig:serverless}, the serverless paradigm generally involves three key stakeholders: developers, cloud providers, and end users\footnote{In some cases, developers can also serve as end users.}. 
Serverless cloud enables developers to simply focus on developing and deploying applications as serverless functions without the burden of managing underlying infrastructure, which significantly enhances ease of use. Additionally, it rightsizes resource allocation across applications and facilitates fine-grained resource sharing, leading to high efficiency.
For end users, serverless cloud provides deployed applications as services, featuring automatic scaling to accommodate dynamic workload demands. 
Users are billed only for actual resource usage at a fine granularity (e.g., 1~ms~\cite{aws_lambda}).
As a result, serverless computing substantially simplifies application deployment and offers high scalability and cost efficiency.  


Owing to these benefits, serverless computing has been widely exploited across various application domains~\cite{pu_shuffling_2019,jonas_occupy_2017,yu_gillis_icdcs,zhang_mark:_2019,ao_sprocket_2018,ali_batch_nodate,jia_nightcore_2021,carver_wukong_2020,zhang_caerus_nodate,lambda_scenarios,sreekanti_cloudburst_2020,sequoia_socc,serverless_schedule_socc}. 
One prominent use case is data analytics, where analytical systems are built and deployed on serverless cloud platforms, providing Query-as-a-Service (QaaS) to end users.
This approach allows data analytics systems to seamlessly leverage the on-demand, elastic resources provided by serverless cloud platforms, enabling high scalability for processing analytical queries that require highly parallel tasks~\cite{pu_shuffling_2019,jonas_occupy_2017,muller_lambada_2020,carver_wukong_2020,zhang_caerus_nodate,li_minflow_2024,jin_ditto_2023}.

However, deploying data-intensive analytics workloads on serverless platforms brings well-known challenges such as data shuffling overhead and inefficient resource allocation, which can significantly impair the end-to-end query performance.
An array of solutions has been proposed to improve the efficiency of serverless analytics, which can be broadly categorized into two approaches: frameworks atop general-purpose serverless platforms and specialized serverless systems. 
Analytical frameworks, such as PyWren~\cite{jonas_occupy_2017} and Nimble~\cite{zhang_caerus_nodate}, directly run atop readily available, general-purpose serverless platforms like AWS Lambda.
This approach typically treats the serverless platform as a black box, relying on external schedulers and storage systems to optimize task execution. 
In contrast, analytics-specialized serverless systems, such as Ditto~\cite{jin_ditto_2023} and MinFlow~\cite{li_minflow_2024}, are tailored for data analytics and re-design the underlying platform for optimized query performance.
While these specialized solutions often deliver superior performance, they are limited to specific workloads and fail to run alongside other applications in a shared environment---a key feature of general-purpose platforms to enhance resource efficiency. 
Moreover, developing such specialized systems from scratch requires significant engineering efforts. 
Such a trade-off---between compatibility, usability, and efficiency---are also observed in many other serverless application domains~\cite{next_serverless_2021}.

Therefore, we pose the following question: \textit{can we mitigate the hard trade-off, by enabling domain-specialized system optimizations for improved performance while retaining a shared, resource-efficient, and easy-to-use serverless platform?}
To this end, we advocate an \textbf{\emph{extensible}} design principle for serverless computing.
Specifically, we propose that the serverless platform should satisfy the three requirements. 
\textbf{\emph{1) Infra-less:}} The platform should enable application developers to easily build specialized frameworks without managing low-level infrastructure details, ensuring high usability. 
\textbf{\emph{2) Generality:}} The platform should be made general enough to support various serverless workloads in a shared environment.
\textbf{\emph{3) Customizability:}} The platform should also enable developers to extend and customize system behaviors for domain-specialized optimizations.

In this paper, we focus on data analytics as a representative use case for serverless computing and study how an extensible approach can be realized~\footnote{Other popular serverless applications, such as video processing~\cite{ao_sprocket_2018,fouladi_encoding_nodate}, graph analytics~\cite{liu_faasgraph_2024}, and machine learning~\cite{yang_infless_2022,faaswap,dilu}, can also benefit from our extensible design, which we discuss in \S\ref{sec:discussion}.}.
We note that system optimizations for serverless analytics can generally fall into the control plane and the data plane. 
The control-plane optimizations, including efficient function scaling and scheduling~\cite{zhang_caerus_nodate,jonas_occupy_2017,li_minflow_2024,jin_ditto_2023}, often rely on both domain and system-level knowledge. 
For instance, achieving optimal Join performance requires judiciously determining the Join approach and the function parallelism based on the data distribution and cluster configurations (details in \S\ref{sec:case}).
In contrast, the data plane optimizations, such as fast data sharing and function startups~\cite{,mahgoub_sonic_nodate,lu_serialization_2024,szekely_unifying_2024,akkus_sand:_2018,pheromone}, are primarily tied to the underlying serverless platform. 
These optimizations are naturally applicable to various application domains including serverless analytics.

\begin{figure}
    \centering
    \includegraphics[width=0.48\textwidth]{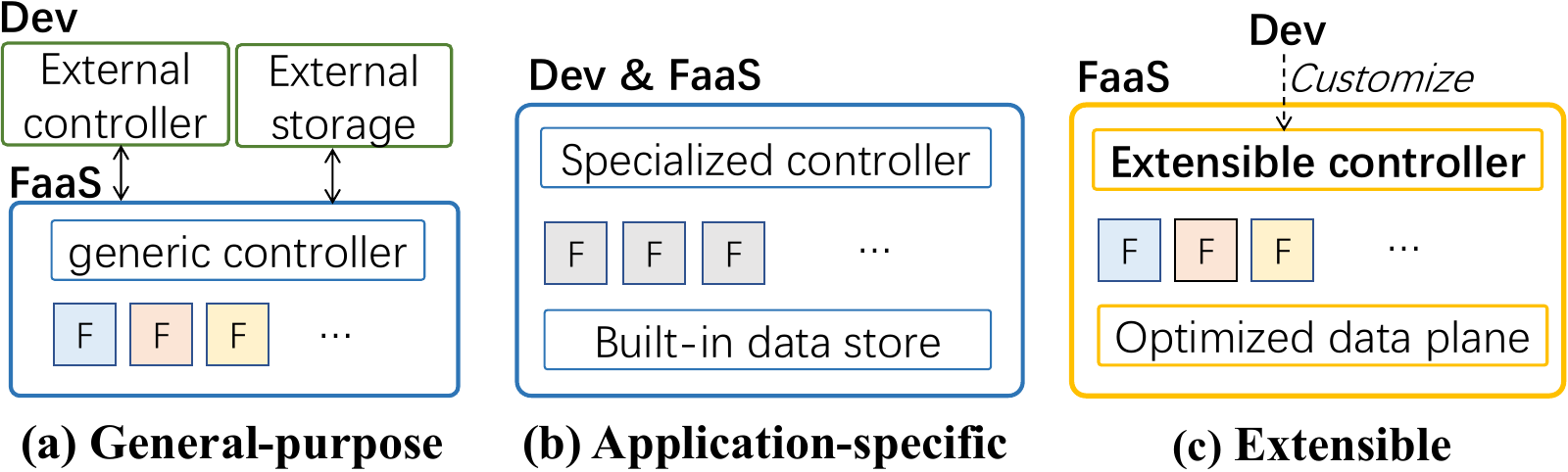}
    \caption{Developing applications atop three kinds of serverless platforms. (a) General-purpose: Developers rely on external components to support complex application needs. (b) Application-specific: Developers redesign FaaS platforms for target applications. (c) Extensible (our vision): Developers simply customize system behaviors for target applications through easy-to-use interfaces.}
    \label{fig:extensible}
    \vspace{-0.2in}
\end{figure}

Based on these observations, we propose to make the control plane of serverless platforms extensible (Fig.~\ref{fig:extensible}). 
The platform provides system-level knowledge to developers, allowing them to customize control-plane behaviors and perform domain-specialized optimizations.
Meanwhile, the platform itself supports an efficient data plane that frees developers from dealing with low-level data sharing and function startup details~\footnote{A rich body of work focuses on data-plane improvements in serverless platforms~\cite{du_catalyzer_2020,ustiugov_benchmarking_2021,ao_faasnap_2022,cadden_seuss_2020,mahgoub_sonic_nodate,lu_serialization_2024,pheromone}, which are complementary to our proposal.}.
This approach effectively meets the desired properties of an extensible serverless platform.
Developers can directly operate on an efficient, shared serverless platform without managing low-level infrastructures (\textbf{\emph{Infra-less}} and \textbf{\emph{Generality}}); they also have the flexibility to extend control-plane system behaviors through easy-to-use interfaces when necessary (\textbf{\emph{Customizability}}).

To realize the extensible design principle, we introduce a prototype of a novel serverless system, \SysName.
\SysName proposes the abstraction of decision workflows with user-friendly interfaces, which allows developers to define custom function scaling and scheduling logic to accommodate application-specific requirements.
\SysName also leverages a decentralized, extensible control plane, which facilitates fine-grained resource sharing across multiple applications. 
We have implemented serverless analytics queries on the \SysName prototype. 
Preliminary results demonstrate that \SysName enhances query execution through custom scheduling strategies and achieves high overall resource efficiency.

We summarize the benefits that the extensible design principle brings to stakeholders in the serverless ecosystem (Fig.~\ref{fig:serverless}).
\textbf{\emph{1) Serverless cloud providers:}} The extensible design enables cloud providers to support a broader range of applications with diverse requirements. 
By accommodating a wider variety of workloads and facilitating efficient resource sharing among them, cloud providers can expand their market reach, improve resource utilization, and enhance profitability.
\textbf{\emph{2) Developers:}} This design alleviates the burden of managing specific system infrastructures, enhancing ease of use compared to building and maintaining specialized serverless systems. 
Additionally, developers benefit from reduced resource costs as the underlying serverless platform supports fine-grained resource sharing across various applications.
\textbf{\emph{3) End users:}} For end users, such as those submitting queries to data analytics systems, the extensible approach delivers performance comparable to that of specialized systems at a lower cost. 
These cost savings are achieved through higher resource utilization and reduced overall platform expenses.

\section{Properties of Serverless}
\label{sec:background}



Serverless computing, emerging as a next-generation cloud computing paradigm, has gained significant traction in recent years. 
It provides a high-level abstraction of serverless functions, liberating developers from the burden of managing underlying infrastructure (hence the term ``serverless''). 
Additionally, serverless computing offers high elasticity that supports automatic resource provisioning in response to workload changes. 
These advantages have driven the widespread adoption of serverless cloud platforms~\cite{aws_lambda,google_cloud_function,azurefunc,ali_fc}. 
We summarize the key properties that define an ideal serverless platform as follows.  

\emph{\textbf{1) Ease of use:}} Serverless computing aims to simplify application development by minimizing the technical complexity in utilizing and managing cloud resources. This ensures a streamlined and developer-friendly experience.

\emph{\textbf{2) Resource efficiency:}} A key goal of serverless computing is to improving resource efficiency. By deploying applications as elastic and lightweight functions, serverless platforms facilitate efficient and fine-grained resource sharing with substantial cost savings. 

\emph{\textbf{3) Satisfactory performance:}} Cloud applications often have diverse performance requirements, such as low latency and Service-Level-Objective (SLO) attainment. It is crucial for serverless platforms to deliver satisfactory performance to meet the varying demands across applications.

\section{Case Study: Serverless Analytics}
\label{sec:case}

While existing serverless platforms can efficiently host simple and stateless workloads, they face significant challenges when deploying real-world complex workloads, such as data analytics~\cite{jonas_occupy_2017,pu_shuffling_2019,muller_lambada_2020,perron_starling_2020}, video processing~\cite{ao_sprocket_2018,fouladi_encoding_nodate}, stream processing~\cite{song_sponge_2023}, and machine learning~\cite{carreira_cirrus_2019,yu_gillis_icdcs}. 
These workloads typically require orchestrating a large number of functions and managing substantial intermediate state exchanges among them. 
However, current serverless platforms often fail to efficiently handle these complexities, leading to violations of their key properties---specifically, limited usability, resource inefficiency, or suboptimal performance (\S\ref{sec:background}).
To illustrate this limitation, we examine serverless data analytics as a case study.

\begin{figure}
    \centering
    \includegraphics[width=0.47\textwidth]{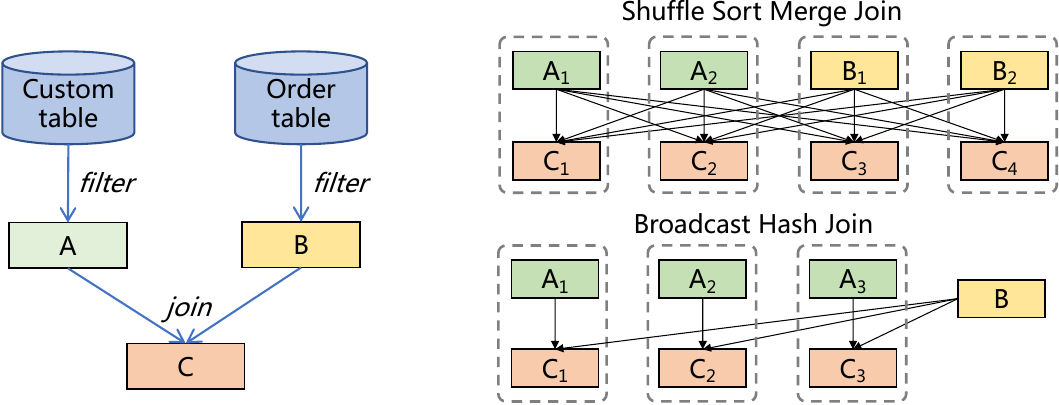}
    \caption{\texttt{Join} exmaple.}
    \label{fig:case}
    \vspace{-0.1in}
\end{figure}

\begin{figure*}
    \centering
    \includegraphics[width=0.98\textwidth]{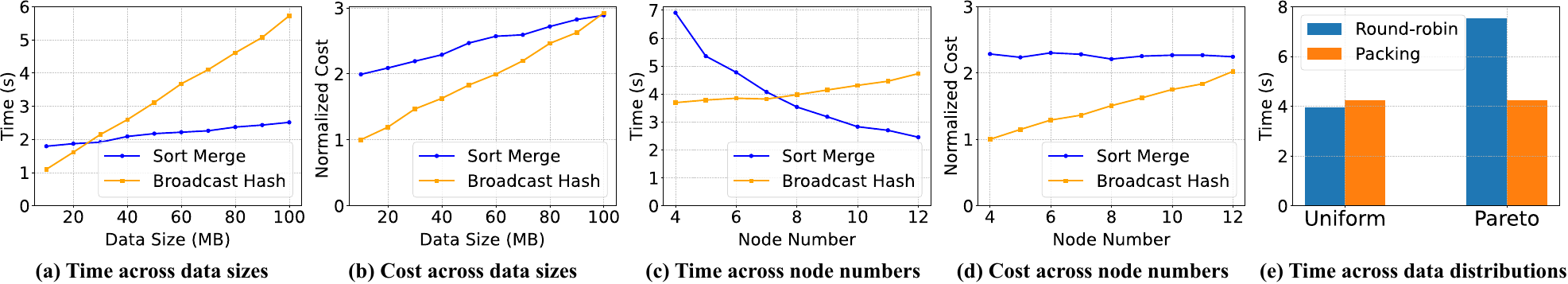}
    \caption{Completion times and normalized costs of \texttt{Join} operation under various cases.}
    \label{fig:join_test}
    \vspace{-0.05in}
\end{figure*}

\PHB{Characteristics of serverless analytics.}
Serverless computing has become an appealing option for analytical workloads that demand launching thousands of parallel tasks for data processing, with tasks running as highly elastic functions.
However, achieving efficient serverless analytics remains a significant challenge. The system must adopt efficient execution plans---scheduling and coordinating parallel tasks effectively---to minimize I/O overhead of intermediate data transfers and optimize query completion times.
Notably, \emph{the optimal configuration of execution plans heavily depends on both system-level knowledge and application semantics.}

We take as an example the Join, a key operation across data analytical queries.
Fig.~\ref{fig:case} shows two different approaches to join the two tables: Sort-Merge Join, which shuffles data records with the same keys to the same nodes for processing, and Hash Join, which broadcasts the entire smaller table to all nodes hosting partitions of the larger table.
We evaluate the performance and cost of these Join implementations under varying scenarios in Fig.~\ref{fig:join_test}.
Fig.\ref{fig:join_test} (a) and (b) present the completion times and normalized cost (measured as resource-time) for joining a 400MB table with another table ranging in size from 10MB to 100MB, using a 12-node cluster (detailed testbed described in \S\ref{sec:eval}).
Hash Join performs better than Sort-Merge Join when the smaller table is under 30~MB but becomes less efficient as the table size increases due to growing broadcasting overhead.
Fig~\ref{fig:join_test} (c) and (d) compare the approaches under varying cluster sizes, with the data sizes of the two tables fixed at 400MB and 80MB, respectively.
As the cluster size grows, Hash Join incurs significant communication overhead, eventually resulting in longer latency compared to Sort-Merge Join. However, Hash Join generally achieves lower costs due to its relatively lighter computational load.
Additionally, Fig.~\ref{fig:join_test} (e) compares the completion times of two common scheduling strategies, round-robin scheduling and packing functions onto few nodes, under different data distributions. 
Using a total of 8 nodes, we evaluate these strategies across uniform and Pareto (skewed) distributions. Round-robin performs well under uniform distribution, while function packing proves more effective for handling skewed data in Pareto distribution scenarios.

To conclude, the optimal configurations of function implementation and scheduling are highly dependent on system-level knowledge, such as data distributions and cluster environments. 
Furthermore, applications often have diverse requirements. For example, some queries may be latency-sensitive, requiring tight completion times, while others may prioritize resource cost savings. 
Therefore, developing such serverless applications requires developers to work with serverless platforms to address these specific requirements. 

\PHB{Existing approaches.}
Existing serverless analytics systems can be generally categorized into two main approaches.

\emph{\underline{1) Frameworks atop general-purpose serverless platforms}} rely external controllers to schedule and coordinate 
functions and leverage external storage systems to manage state sharing across them, as shown in Fig.~\ref{fig:extensible} (a). 
For example, PyWren~\cite{jonas_occupy_2017}, Lambada~\cite{muller_lambada_2020}, Wukong~\cite{carver_wukong_2020}, and Nimble~\cite{zhang_caerus_nodate} employ external schedulers to optimize task execution atop AWS Lambda; Locus and Pocket~\cite{klimovic_pocket:_2018} introduce multi-tier storage solutions for efficient data sharing among serverless analytics functions.
These frameworks are flexible and easily portable to general-purpose serverless platforms. 
However, they treat serverless platforms as black-box systems, lacking internal visibility into system knowledge and execution status. This limitation makes it challenging to implement fine-grained optimizations, \emph{thereby resulting in suboptimal performance for complex workloads---a violation of the key properties in \S\ref{sec:background}.} 

\emph{\underline{2) Application-specific serverless systems}} open the black box of serverless platforms by redesigning the underlying platform, such as internal data storage and specialized controllers, to enhance performance for target applications (Fig.~\ref{fig:extensible} (b)).
For example, Ditto~\cite{jin_ditto_2023} and Minflow~\cite{li_minflow_2024} integrate built-in data storage to improve state sharing and design efficient task execution plans to enhance data locality for analytics queries. These optimizations are highly effective in improving performance for analytical workloads.
However, these systems are typically tailored to the narrow application domain, requiring significant engineering effort to design specialized serverless systems for each use case. 
Specifically, developers are forced to build specialized systems from scratch, reintroducing complexities such as resource management and scaling, which diminishes usability. Furthermore, these systems are unable to co-execute alongside a broader range of applications, limiting the ability to improve resource utilization. 
Consequently, \emph{this approach fails to achieve high usability and resource efficiency as outlined in §\ref{sec:background}.}

\PHB{Rethinking serverless solutions.}
The dilemma above is not unique to serverless analytics but has been observed across many other domains~\cite{liu_faasgraph_2024,next_serverless_2021,yang_infless_2022,fouladi_encoding_nodate,fouladi_laptop_2019,mahgoub_sonic_nodate}. Developers often face a challenging tradeoff: building solutions atop general-purpose serverless platforms leads to suboptimal performance for complex applications, while developing application-specific systems sacrifices the ease of use and fine-grained resource sharing that make serverless computing appealing.
This raises a critical open question: \emph{Can a single serverless platform incorporate domain-specialized optimizations to enhance performance for particular applications, while simultaneously supporting diverse workloads in a shared, resource-efficient environment without compromising usability, i.e., effectively meeting all the key properties in \S\ref{sec:background}?}

\section{Our Vision: Extensible FaaS}
\label{sec:vision}

To address the aforementioned question, we advocate for an \emph{\textbf{extensible}} design principle in serverless computing. 
To meet the key properties in \S\ref{sec:background}, a desired serverless platform should follow these principles. 

\emph{\textbf{1) Infra-less:}} The platform should free developers from managing low-level infrastructure details, ensuring high usability by abstracting away resource management complexities.

\emph{\textbf{2) Generality:}} To achieve resource efficiency, the platform must broadly support diverse applications in a shared environment, which provides three main benefits. First, sharing a large number of jobs stabilizes overall platform resource utilization and improves efficiency~\cite{zhang_shepherd_nodate,ali_eurosys23}. Second, different applications have varying resource requirements (e.g., compute- or memory-intensive). Collocating these tasks further enhances resource utilization~\cite{data_facebook,serverless_schedule_socc}. Third, applications often differ in their performance needs. Combining latency-sensitive online tasks with offline tasks enables efficient resource sharing without compromising performance~\cite{sahraei_xfaas_2023,azure_asplos25,sequoia_socc}. 

\emph{\textbf{3) Customizability:}} The platform must allow developers to customize system behavior, which is essential for an extensible serverless platform.
As discussed in \S\ref{sec:case}, achieving optimal performance requires combining application- and system-level knowledge. Developers should have access to system insights and be able to incorporate application-specific semantics into the platform to ensure satisfactory performance.

To implement these extensible principles, we have a key insight that guides system design. 
 We note that application-specific optimizations, such as resource scaling and function scheduling, are primarily carried out at the control plane of serverless platforms. 
 In contrast, data-plane behaviors---such as data sharing and function startups---operate at a lower level and are largely independent of application semantics.
Therefore, \emph{a desired platform should offer an extensible control plane that enables the incorporation of application-specific optimizations while maintaining an efficient, general-purpose data plane shared across workloads}, as shown in Fig.~\ref{fig:extensible} (c). 
Since data-plane optimizations have been extensively studied in serverless computing research~\cite{du_catalyzer_2020,ustiugov_benchmarking_2021,ao_faasnap_2022,cadden_seuss_2020,mahgoub_sonic_nodate,lu_serialization_2024,wang_faasnet_nodate,pheromone,kuchler_function_2023,li_dataflower_2023,pheromone-ton}, these techniques can be integrated into the platform and complement application-aware control-plane strategies to achieve efficient execution across applications.

To conclude, an extensible platform should expose system-level knowledge to developers, enabling them to extend control-plane mechanisms with application-specific semantics and requirements. This approach effectively fulfills the three key properties of serverless computing (\S\ref{sec:background}).
First, it ensures high usability and alleviates developers from manually managing infrastructure details (i.e., \textbf{\emph{Ease of use}}).
Moreover, a unified, general-purpose serverless platform can broadly support diverse workloads and facilitate fine-grained resource sharing (i.e., \textbf{\emph{Resource efficiency}}).
Finally, developers can customize the system behavior to accommodate domain-specialized performance optimizations (i.e., \textbf{\emph{Satisfactory performance}}).

\section{\SysName Prototype}
\label{sec:design}

\begin{figure}
    \centering
    \includegraphics[width=0.45\textwidth]{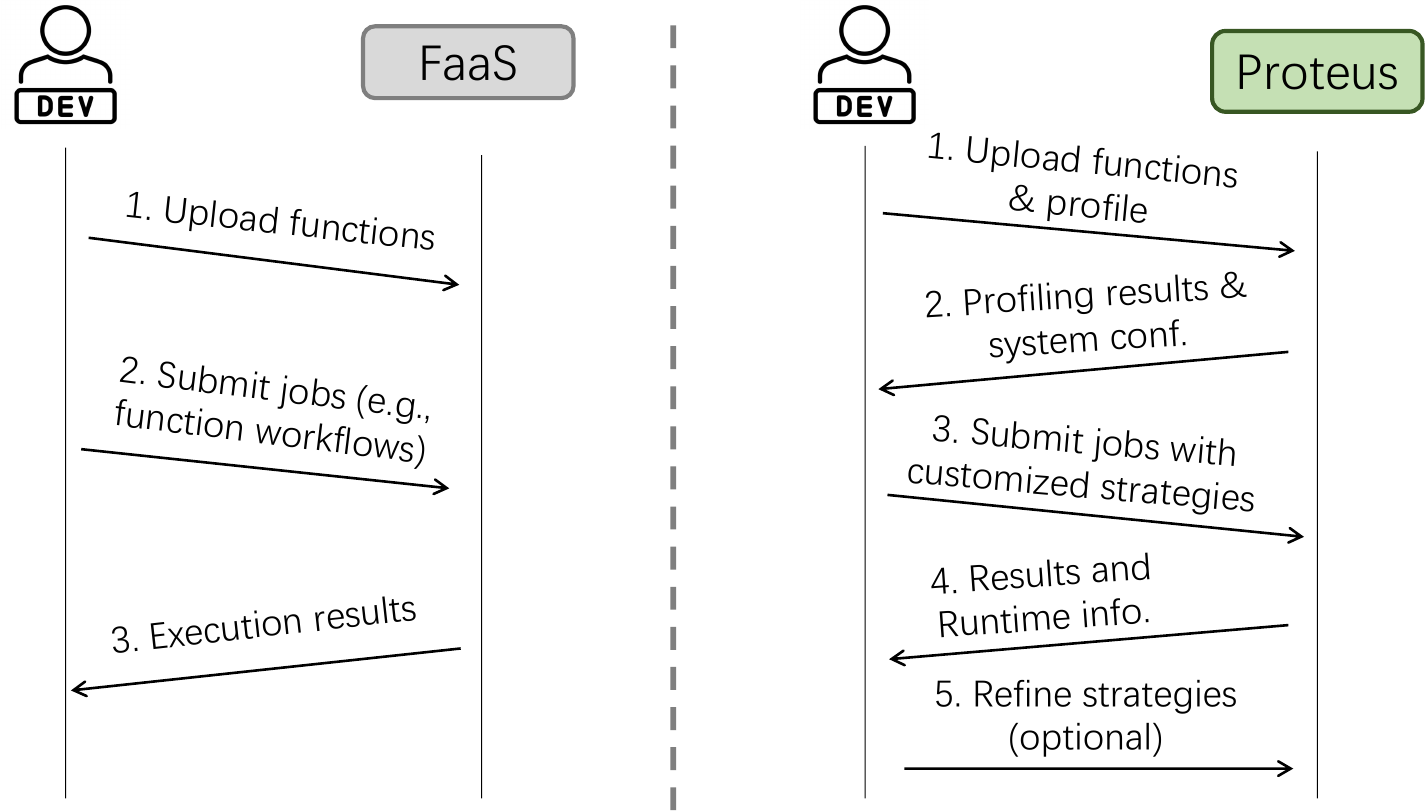}
    \caption{Comparison between existing general-purpose serverless platforms (left) and \SysName (right).}
    \label{fig:overview}
    \vspace{-0.1in}
\end{figure}

We propose \SysName, a prototype of a serverless computing platform that realizes the extensible design principle.
Fig.~\ref{fig:overview} illustrates the workflow for deploying and executing applications on existing general-purpose serverless platforms and \SysName~\footnote{FaaS platforms support various function triggering and execution methods, which we show in simplified form here.}. 
Compared with existing platforms, \SysName allows developers to customize the control-plane behaviors to accommodate specific requirements for their applications.
It provides system configurations and profiling results (Step 2), which developers can combine with application-level semantics to build customized optimization strategies.
After completing the execution, \SysName also returns the runtime information in addition to the results (Step 4), allowing developers to further refine their optimization strategies for subsequent runs.

The remainder of this section describes the key design of \SysName, including its programming model and extensible control plane.
While data-plane designs and optimizations are not the focus of this work, \SysName can integrate existing solutions from a rich body of related research~\cite{du_catalyzer_2020,ustiugov_benchmarking_2021,ao_faasnap_2022,cadden_seuss_2020,mahgoub_sonic_nodate,lu_serialization_2024,wang_faasnet_nodate,pheromone,kuchler_function_2023,shillaker_faasm_2020}.
We will also discuss several key questions and future directions for \SysName in \S\ref{sec:discussion}.

\subsection{Programming Model}
\label{sec:programming} 

\SysName enables developers to customize the scaling and scheduling logic of their functions.
A key challenge lies in providing a programming model that simplifies control-plane decision making and performs these decisions at runtime. 
To address this, \SysName introduces the abstractions---decision nodes and workflows---along with easy-to-use interfaces.

\PHB{Decision nodes and workflows.} 
In the control plane, \SysName abstracts the scheduling of a group of (one or more) functions as a ``decision node'', and represents an application (e.g., an analytical query) as a workflow of these nodes, i.e., ``decision workflow''.
Each decision node determines the scaling and scheduling of downstream functions based on runtime information, including available resources and data distribution across the cluster. 
Developers can customize the logic in each decision node and define end-to-end decision workflows. 
This design allows them to embed application semantics into system optimizations and dynamically adjust control-plane strategies in response to changing runtime conditions.
We note that decision workflows can enhance the function workflow used in many serverless platforms.
While function workflows merely define the function execution order, decision workflows further empower developers to specify the scaling and scheduling of these functions at runtime.
For simple applications that do not require control-plane customizations, \SysName provides default system mechanisms and falls back to standard function workflows.

\begin{figure}[tb]
\begin{lstlisting}
input data_dist, node_status
output decision_tuple (func, scale, schedule)

# Get data sizes of table A and B
sizeA, sizeB = data_dist.A.size, data_dist.B.size 
# Get nodes containing table A and B
nodeA, nodeB = data_dist.A.loc, data_dist.B.loc
# T1, T2 are predefined thresholds
if sizeA / sizeB < T1 and |nodeA| > T2: 
    func = "merge_join"
    scale = (sizeA + sizeB) / a # proportional to size
    schedule = ("round-robin", nodeA U nodeB)
else:
    func = "hash_join"
    scale = num_of_avail_slots(node_status, nodeA) 
    schedule = ("packing", nodeA) 
\end{lstlisting}
\vspace{-.4in}
\caption{Example of a Join decision node.}
\label{fig:api_example}
\vspace{-.2in}
\end{figure}

\PHB{APIs.}
\SysName provides an interface that enables developers to embed custom scheduling and scaling logic within decision nodes. Fig.\ref{fig:api_example} demonstrates an example decision node for a Join operation (Fig.\ref{fig:case}).
This interface takes as input the data distribution and the current node status, including the local data volume for each table and the available function slots on each worker node. It produces a control decision tuple, specifying the function to invoke, the number of function instances to scale, and the scheduling policy to apply.
In this example, the decision node analyzes the data distribution and cluster size to decide whether to invoke a merge-join or hash-join function (line 9). 
Based on observations in Fig.~\ref{fig:join_test}, merge-join performs better in large clusters and when data volumes across tables are balanced. Developers can leverage such application-level characteristics to optimize decisions for improved performance. 
For each scenario, the decision node determines the appropriate number of function instances and their scheduling onto selected nodes (lines 10-12, 14-16). 
Scheduling strategies, such as round-robin or packing, are specified along with a set of cluster nodes. With such a decision output, \SysName can generate a detailed function execution plan accordingly.
These decision nodes are executed at runtime as part of the end-to-end workflow, enabling dynamic, application-specific optimizations during execution.

\subsection{Extensible Control Plane}

\SysName's control plane monitors runtime status (e.g., application data, cluster nodes, and function instances) and performs function scheduling and resource management accordingly.
To achieve the extensibility, \SysName adopts a decentralized design in its control plane.
The control plane consists of a global controller and logically private controllers for each application. The global controller coordinates resource allocations across applications and maintains a comprehensive view of system resources, offering (all or parts of) available resources to private controllers. Each private controller, in turn, manages application-specific execution by tracking application-level information (e.g., decision workflows and intermediate data distribution) and executing custom scheduling logic embedded within decision nodes.

For resource sharing across applications, \SysName can employ an Omega-like approach~\cite{schwarzkopf_omega_2013}, enabling private controllers to make application-level scheduling decisions while leveraging global resource status. In multi-tenant scenarios, where applications compete for shared resources, the global controller resolves conflicts by prioritizing higher-priority tasks.
Such prioritization is particularly effective in real-world FaaS platforms, where low-priority, delay-tolerant workloads are common~\cite{sahraei_xfaas_2023}. By prioritizing tasks, \SysName can significantly enhance overall resource efficiency without compromising the performance requirements of high-priority applications (see preliminary results in \S\ref{sec:eval}).

\section{Preliminary Results}
\label{sec:eval}

\begin{figure}
    \centering
    \includegraphics[width=0.41\textwidth]{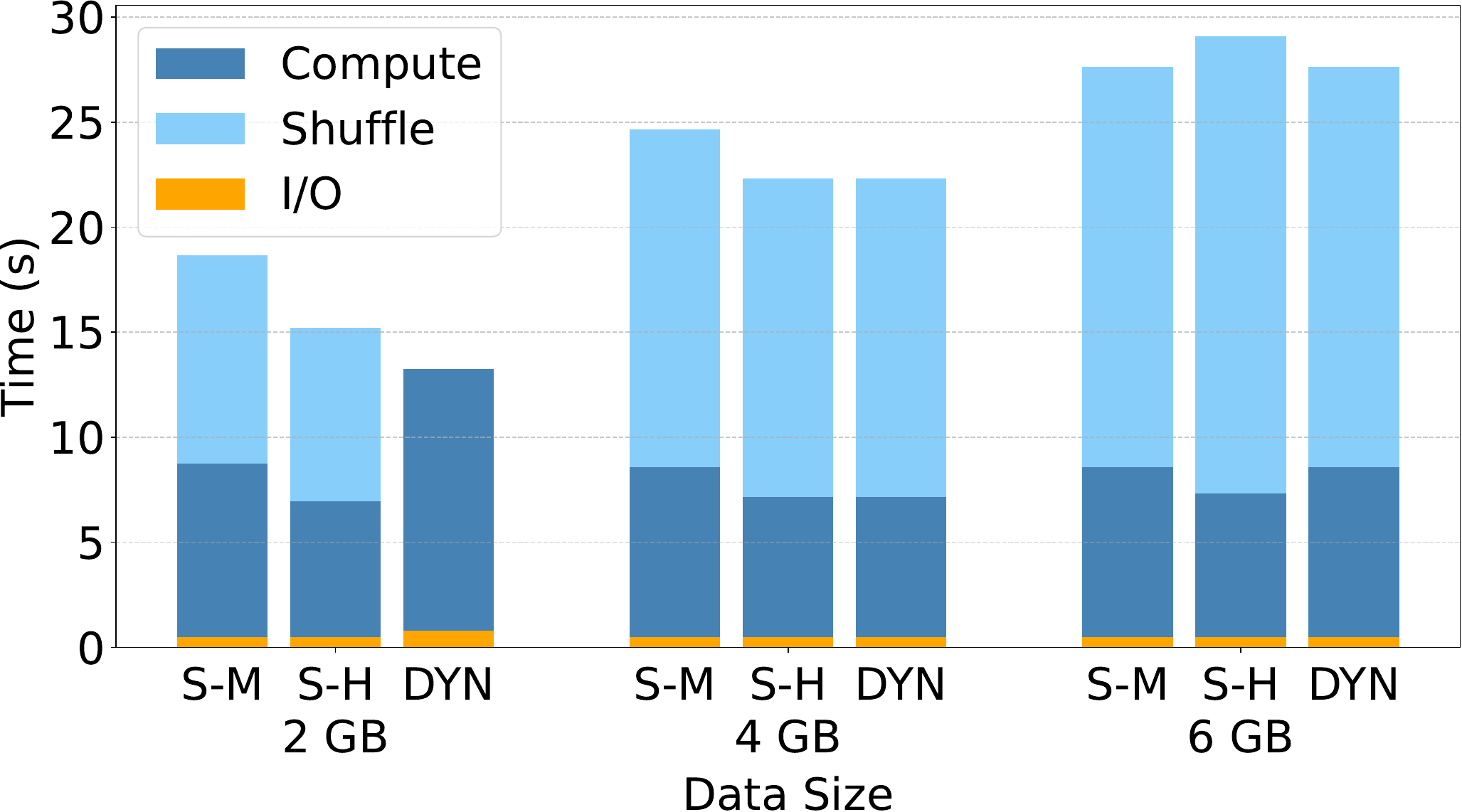}
    \caption{Query performance under various strategies.}
    \vspace{-0.1in}
    \label{fig:end_to_end}
    \vspace{-0in}
\end{figure}

We have built a preliminary prototype of \SysName using 4k lines of C++ code and implemented common analytics operators as \SysName's functions using 1k lines of code.
The current prototype focuses on incorporating custom control-plane logics into the system.
Our ongoing work concentrates on optimizing strategies from the developer side to further enhance data analytics performance and expanding \SysName's extensible control plane for fine-grained resource sharing among a broad range of applications (more details in \S\ref{sec:discussion}).
In the data plane, \SysName executes functions as containers and employs shared-memory data exchange, following the practice of existing platforms~\cite{pheromone,jia_nightcore_2021,shillaker_faasm_2020}.

We evaluate \SysName using a sub-query from TPC-DS, which consists of two MapReduce phases and a Join phase.
We deploy \SysName on a 6-node cluster of \texttt{c5.2xlarge} instances on AWS EC2, with each instance having 8 vCPUs and 16 GB of memory. 

\PHB{Query completion time.}
We first evaluate the query performance under \SysName.
We implement three strategies using \SysName's APIs, Static Merge Join (S-M), Static Hash Join (S-H), and Dynamic (DYN).
The two static strategies employ fixed join operators and scale function instances proportionally to data sizes, while DYN supports dynamic configurations and function scheduling, such as the Join decision node in Fig.~\ref{fig:api_example}.
Fig.~\ref{fig:end_to_end} depicts the end-to-end query latency under varying data sizes.
Among these strategies, DYN outperforms the others due to its adaptability to different cases: for 2~GB data, DYN consolidates all functions on a single node to eliminate shuffle overhead (i.e., function packing); for 4~GB and 6~GB data, DYN dynamically switches between hash join and merge join for improved overall performance.
This result demonstrates the ability of \SysName to enable control-plane optimizations from the developers, which can effectively enhance query performance across diverse cases.

\begin{figure}
    \centering
    \includegraphics[width=0.41\textwidth]{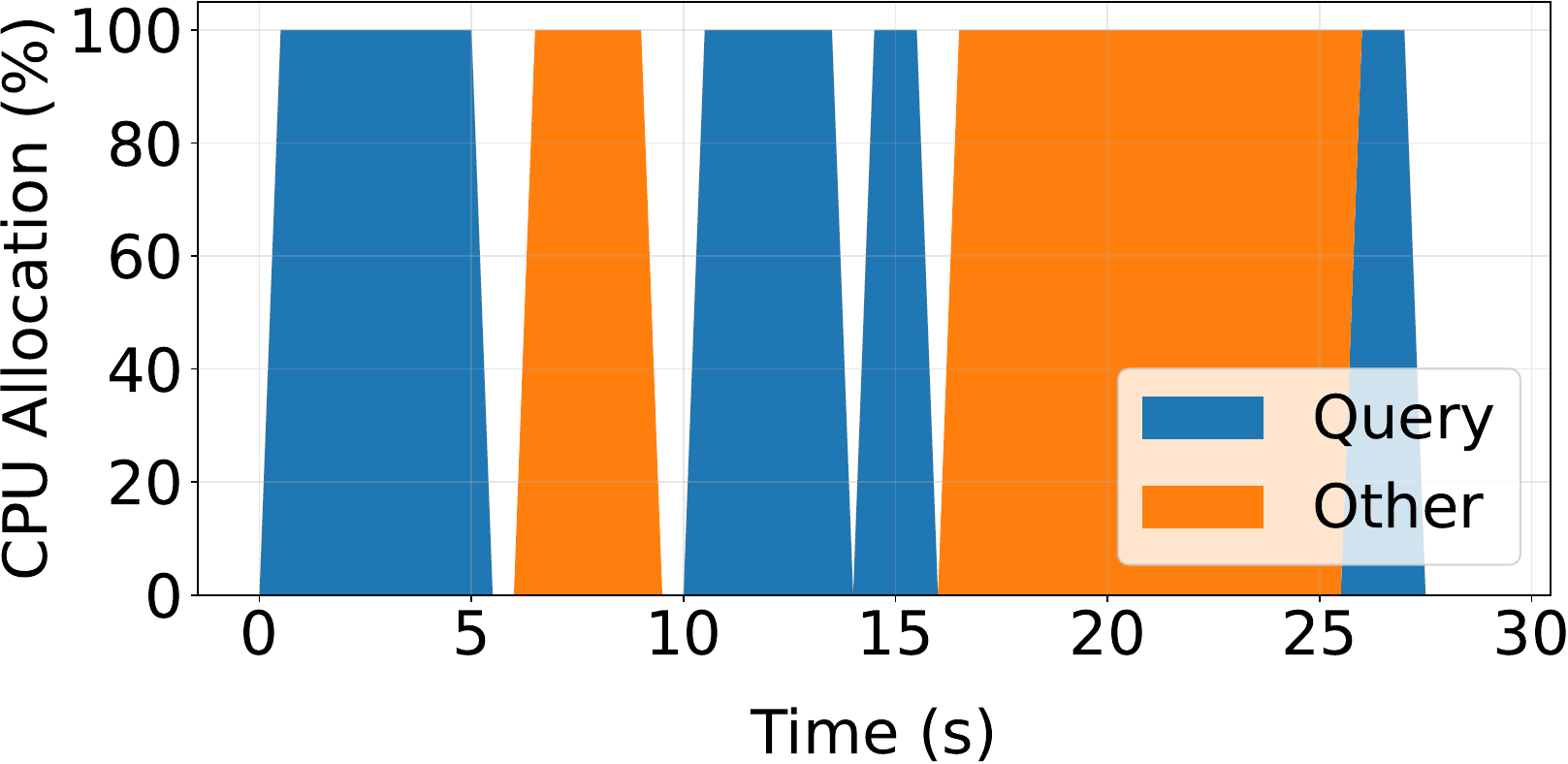}
    \caption{Resource allocation rates of \SysName.}
    \label{fig:resource}
    \vspace{-0.2in}
\end{figure}

\PHB{Fine-grained resource sharing.}
We further evaluate resource efficiency of \SysName when the query job co-runs with other applications.
In this experiment, we deploy a set of simple function chains as low-priority, delay-tolerant tasks, following the production practices from Meta’s XFaaS platform~\cite{sahraei_xfaas_2023}. 
These background tasks execute when resources occupied by the query job become available.
Fig.~\ref{fig:resource} shows CPU allocation rates of \SysName when co-running the query job (with 6~GB input data) alongside low-priority background functions. 
During the query execution, we observe two shuffle periods (5-10s and 16-26s) characterized by intensive data communications, where the query job imposes minimal CPU demand. 
As a result, \SysName efficiently schedules background functions to utilize the available resources, achieving high CPU allocation rates and improving overall resource efficiency. 
This demonstrates \SysName's ability to provide generality and support fine-grained resource sharing across diverse applications, maximizing resource utilization without compromising the performance of critical workloads (e.g., the analytical queries).

\section{Discussion and Future Work}
\label{sec:discussion}

We discuss several key questions and future directions for \SysName.

\Q{1) Can the increased complexity of \SysName be justified?}
While \SysName introduces additional complexity over existing serverless platforms (Fig.~\ref{fig:overview}), this design is specifically for addressing large, complex workloads with demanding performance requirements---a scenario typically handled by experienced developers.
Compared to building a specialized serverless system from scratch, \SysName offers a more accessible solution while delivering significant cost savings through fine-grained resource sharing across applications. 
For simpler workloads without customization needs, \SysName defaults to standard mechanisms and operates similarly to other FaaS platforms, ensuring ease of use.

\Q{2) Can \SysName support heterogeneous hardware?}
As serverless platforms increasingly adopt heterogeneous hardware~\cite{fu_serverlessllm_2024,faaswap,heter_serverless_2022}, improving resource utilization and reducing data transmission overhead across diverse hardware have become critical challenges~\cite{dilu,fuyao_2024,wu2024faastube}. 
With data-plane support for heterogeneous hardware, such as enabling efficient data transfer among heterogeneous functions, \SysName can be applied in these environments and enhance performance through control-plane optimizations. 
We leave it as a future work.

\Q{3) How can \SysName be applied in real-world FaaS?}
Mainstream FaaS platforms often decouple the control plane from the data plane~\cite{wang_faasnet_nodate,sahraei_xfaas_2023}, e.g., separating function scheduling from function sandbox optimizations.
This clear separation creates opportunities to implement \SysName's extensible design in practice.
With its focus on an extensible control plane, \SysName can be integrated into existing serverless platforms while retaining their underlying infrastructure and data-plane optimizations. 
We will explore the integration as a future work.

\Q{4) Can \SysName benefit other application domains?}
\SysName's extensible approach is not limited to data analytics and can be applied to other applications that benefit from control-plane customizations to meet specific requirements. Many popular serverless applications, such as video processing~\cite{fouladi_encoding_nodate,ao_sprocket_2018}, graph analytics~\cite{liu_faasgraph_2024}, and machine learning~\cite{faaswap,dilu,ali_batch_nodate}, require complex execution plans to achieve optimal performance.
Applying \SysName's extensible approach to these scenarios can improve the usability and efficiency.

Machine learning inference is a prevalent serverless use case that can demonstrate the benefits offered by \SysName. A key challenge in serverless inference systems is to efficiently allocate resources and schedule request batches to balance inference performance and resource cost~\cite{ali_optimizing_2022,yang_infless_2022,ali_batch_nodate,yu2025lambdascale}. For instance, batching a large number of requests improves resource utilization but increases completion times, whereas smaller batches reduce latency but may compromise efficiency.
\SysName's extensible approach allows developers to easily customize the control-plane strategies, such as optimized resource allocation and scheduling, based on the specific characteristics of machine learning models and their performance requirements.
This enables developers to achieve improved performance without the need to build a specialized serverless system from scratch.
Extending \SysName to other application domains will be a focus of our future work.

\section{Conclusion}


This paper introduces an extensible design principle for serverless computing that enables domain-specialized optimizations in a shared, easy-to-use serverless platform. 
We prototype this approach in \SysName, a serverless platform that provides a novel abstraction of decision workflows, allowing developers to customize control-plane logics for their particular applications.
Preliminary results show that \SysName effectively improves performance for data analytical queries and supports fine-grained resource sharing across various applications.

\bibliographystyle{plain}
\bibliography{references}

\end{document}